\documentclass[sigconf]{acmart}
\setcopyright{none}
\renewcommand\footnotetextcopyrightpermission[1]{}
\acmConference[IAFM '26]{Workshop on Interpretability and Alignment of Financial Models}{November 2026}{Milan, Italy}
\acmYear{2026}
\title{Certifying Reward Alignment in Portfolio Reinforcement Learning}
\author{Matin Tavakoli}
\affiliation{\institution{Amirkabir University of Technology}\country{}}
\author{Seyedreza Tavakoli}
\affiliation{\institution{Dataism Laboratory for Quantitative Finance}\country{}}
\author{Faezeh Salmani}
\affiliation{\institution{Allameh Tabataba'i University}\country{}}
\author{Amir Amini}
\affiliation{\department{}\institution{Princeton University}\country{}}
\affiliation{\institution{Dataism Laboratory for Quantitative Finance}\country{}}
\author{Mohammad Hosein Rohban}
\affiliation{\institution{Sharif University of Technology}\country{}}
\author{Gholamali Aminian}
\affiliation{\institution{The Alan Turing Institute}\country{}}
\author{Ali Habibnia}
\authornotemark[2]
\affiliation{\institution{Virginia Tech}\country{}}
\affiliation{\institution{Dataism Laboratory for Quantitative Finance}\country{}}
\renewcommand{\shortauthors}{Tavakoli et al.}
\newcommand{\E}{\mathbb{E}}
\newcommand{\Var}{\operatorname{Var}}
\newcommand{\Cov}{\operatorname{Cov}}
\newcommand{\spn}{\operatorname{span}}
\newcommand{\eq}{\mathrm{eq}}

\newtheorem{theorem}{Theorem}

\newcommand{\DistanceReduction}{59.2}
\newcommand{\DeviationReduction}{81.4}
\newcommand{\CentreReduction}{0.57}
\newcommand{\PpoText}{With PPO, the continuation correction reduces mean distance by 33.3\% and maximum deviation gain by 49.8\%. The paired distance difference is $-0.575$ (95\% seed-bootstrap interval $[-0.739,-0.404]$).}

\begin{document}
\begin{abstract}
Financial reinforcement learning often translates an investment objective into an additive risk-adjusted reward. An agent can optimise that reward accurately yet implement a different financial preference. We develop a decision audit for terminal mean--variance equilibrium that separates reward misspecification from learning error. In a finite exogenous-state market, we bound the gain from changing one action using actor, reward and continuation-critic residuals. The certificate ignores state-constant critic offsets, accommodates convex action constraints, and has an independent-sample extension that does not require the equilibrium policy. A continuation-aware actor--critic method estimates the future-policy correction from samples. Across nine controlled markets and twenty seeds, it reduces policy distance by \DistanceReduction\% and one-step deviation gain by \DeviationReduction\% relative to a matched myopic-reward network. A separate PPO comparison tests the same reward intervention within a standard optimiser. Historical tests cross seven covariance estimators and 10, 30 and 49 US industry portfolios over 432 months: joint reward calibration preserves each fitted mean--variance target, but does not improve realised investment performance. Together, these results provide a way to quantify whether a learned decision implements its stated financial objective, separately from how well that objective performs in a backtest.
\end{abstract}
\ccsdesc[500]{Computing methodologies~Reinforcement learning}
\ccsdesc[300]{Applied computing~Economics}
\keywords{financial reinforcement learning, reward alignment, equilibrium, actor--critic, model validation}
\maketitle
\fancypagestyle{acceptedfirstpage}{%
  \fancyhf{}%
  \renewcommand{\headrulewidth}{0pt}%
  \renewcommand{\footrulewidth}{0pt}%
  \fancyfoot[L]{\parbox[t]{\columnwidth}{\footnotesize\raggedright Accepted at the Workshop on Interpretability and Alignment\\of Financial Models (IAFM), ACM ICAIF 2026.\\[3pt]Code, data and experiment outputs are available at\\\url{https://github.com/DataismLab/Reward-alignment}.}}%
}
\thispagestyle{acceptedfirstpage}
\begingroup
\renewcommand{\thefootnote}{\fnsymbol{footnote}}
\footnotetext[2]{Corresponding author: \href{mailto:habibnia@vt.edu}{habibnia@vt.edu}}
\endgroup

\section{Introduction}
An investment mandate specifies preferences over return, risk and costs; an RL agent maximises a reward. The translation can change decisions: replacing variance by a second moment penalises squared expected return; ignoring dependence removes a continuation hedge; correcting risk without scaling costs changes turnover preferences. Accurate reward optimisation or a favourable backtest does not establish mandate alignment.

Kolm and Ritter~\cite{kolm2025}, pp.~90--91, place mandate-aligned reward specification before training. We ask whether the reward induces the intended decisions and the agent implements them. Our target is conditional terminal mean--variance equilibrium. When computable, it permits direct policy comparison; otherwise, we bound the improvement from changing one action while retaining the agent's future policy.

Our residual certificate separates actor, reward and continuation errors. A zero actor gradient can conceal a wrong critic, while large state-constant critic errors can be harmless. Tracking state differences and their propagation distinguishes these cases and supplies an acceptance criterion under explicit assumptions. A continuation-aware learner estimates the correction from samples, without an equilibrium oracle in training.

Controlled actor--critic and PPO comparisons hold optimiser and data access fixed. Historical tests separately vary covariance and rewards: Markowitz~\cite{markowitz1952} distinguishes beliefs from decisions, while DeMiguel et al.~\cite{demiguel2009} show that estimation error can overwhelm diversification gains. Preserving a fitted objective need not improve realised returns.

\paragraph{Contributions.}
We contribute: (i) a certificate for finite-iterate equilibrium deviations, invariant to state-constant critic offsets and extendable to convex constraints and independent data; (ii) continuation-aware sample learning tested in matched neural and PPO comparisons; and (iii) a historical comparison separating alignment, estimation risk and performance. The new link is from reward and learning residuals to financial decision quality, building on established equilibrium learning, mean auxiliaries and concentration bounds.

\paragraph{Roadmap.}
Section~3 defines two distinct targets: terminal equilibrium and fitted one-period allocation. Sections~4--5 certify and learn the former. Section~6 tests sequential learning, then historical one-period calibration; Section~7 interprets validation decisions. Sections~8--10 provide technical and reproduction details.

\section{Related work}
\paragraph{Financial objectives and estimation.}
DeMiguel, Garlappi and Uppal~\cite{demiguel2009}, Tables~3--5, compare fourteen models with $1/N $. Proposition~1 and Figure~1 relate estimation requirements to dimension and opportunities, not a minimum asset count. Table~2 includes ten industries plus the market; we also test finer classifications. We implement linear shrinkage~\cite{ledoit2004identity,ledoit2004correlation}, observed-factor covariance~\cite{fan2008} and QuEST nonlinear shrinkage. Ledoit and Wolf~\cite{ledoit2017}, Corollary~1 and Eqs.~(14)--(15), correct eigenvalues individually while retaining sample eigenvectors. Their Table~8 motivates holding long-only constraints and costs fixed.

G\^arleanu and Pedersen~\cite{garleanu2013}, Eq.~(4) and Propositions~2--3, derive cost-aware dynamic allocation; DeMiguel et al.~\cite{demiguel2024}, Eq.~(6) and Tables~II--III, study conditional multifactor portfolios with costs. Their dynamic targets differ from our rolling one-period reference. Changing covariance or intentionally changing preferences is not reward misspecification.

\paragraph{Financial RL and reward design.}
Kolm and Ritter's earlier hedging work~\cite{kolm2019}, Eqs.~(8)--(10), explicitly treats quadratic increment rewards as an approximation. Jiang et al.~\cite{jiang2025}, Eq.~(8), combine log return and variance; their 2024 formulation~\cite{jiang2024}, Eq.~(4), also considers costs. Habibnia and Soltanzadeh~\cite{habibnia2024}, Eq.~(8), use asymmetric PnL with SAC, shorting and lending. Table~\ref{tab:historical} adapts reward formulations to common states; it does not reproduce these papers' complete agents or backtests.

Objective-aware learning is established. The reward in Simon et al.~\cite{simon2026}, Eq.~(17), is centred mean--variance utility. Gollart and Okhrin~\cite{gollart2025}, Section~4.2.1, compare learned policies with an analytical benchmark. Cong et al.~\cite{cong2026}, Eqs.~(19)--(20), permit differentiable trajectory objectives. Kolm and Ritter~\cite{kolm2025} survey the area and discuss validation against known policies (p.~87), so benchmark comparison itself is not new.

\paragraph{Equilibrium learning and certification.}
Precommitment~\cite{ling2000} and equilibrium~\cite{basak2010} solve different decision problems. Continuation hedges~\cite{czichowsky2013}, equivalent objectives~\cite{bkm2016} and equilibrium RL~\cite{dai2023} precede our work. Lesmana and Pun~\cite{lesmana2025} develop subgame-perfect equilibrium RL and backward policy iteration; our implementation specialises these principles rather than replicating their algorithm. Guo et al.~\cite{guo2026}, Algorithms~1--3, already combine auxiliary rewards, deterministic policy gradients and fixed-point critic learning. We do not propose a new general equilibrium solver.

Sangadi et al.~\cite{sangadi2025}, Theorem~4.3, provide finite-sample actor--critic stationarity analysis for a discounted return mean--variance objective. Our target is instead a posteriori, statewise one-step deviation gain for a given finite-horizon policy. Seminorm learning and irrelevant critic offsets are also established~\cite{chen2025seminorm}. We use that geometry to obtain a financial error certificate. The scalar reward-calibration identity below follows the mean auxiliary in MVPI~\cite{zhang2021}; the empirical Bernstein allowance follows Maurer and Pontil~\cite{maurer2009}. The contribution lies in the resulting decision audit, not in inventing these ingredients.

\section{Objectives and reward calibration}
\subsection{Terminal and fitted references}
Let $Z_t $ be a finite exogenous Markov state and let excess returns $R_t\in\mathbb R^d $ be jointly distributed with $Z_{t+1}$ conditional on $Z_t $. Wealth obeys
\begin{equation}\label{eq:wealth}
 X_{t+1}=sX_t+R_t^\top u_t(Z_t),\qquad H_t=s^{T-t-1},
\end{equation}
where $s>0$ is deterministic. The financial criterion at each state is conditional terminal mean minus $\varphi>0$ times terminal variance. We consider deterministic wealth-independent policies and their one-step deviations, holding future policies fixed. This equilibrium criterion differs from maximising initial-date precommitment value.

For independent returns with mean $m_t $ and covariance $\Sigma_t\succ0$, equilibrium is $u_t^\eq=\Sigma_t^{-1}m_t/(2\varphi H_t)$. With $M_t=\Sigma_t+m_tm_t^\top $ and $\nu_t=m_t^\top M_t^{-1}m_t $, the additive reward
\begin{equation}\label{eq:independent}
 Y_t-\varphi(1-\nu_t)Y_t^2,\qquad Y_t=H_tR_t^\top u_t,
\end{equation}
recovers that policy. This specialises prior equivalent-objective constructions~\cite{bkm2016}. Under dependence, continuation covariance is essential; Eq.~\eqref{eq:independent} alone is insufficient.

For the raw family $\Delta X-\kappa(\Delta X)^2$, even optimal constant scalar tuning leaves a discrepancy. With independent constant moments and $m\ne0$, $d=M^{-1}m $, $\nu=m^\top d $, $b=[2\varphi(1-\nu)]^{-1}$ and fixed $T\ge2$,
\begin{align}\label{eq:raw}
\min_{\kappa>0}D&=c_T|s-1|+O((s-1)^2),\\
c_T&=\|d\|b\sqrt{\tfrac{T(T^2-1)}3+\tfrac{\nu(1-\nu)T(T-1)}2}.\nonumber
\end{align}
Here $D^2=\sum_t\E_{\pi^\kappa}\|\pi^\kappa(t,X_t)-u_t^\eq\|^2$ uses the raw-reward optimum's occupancy. Section~\ref{app:raw} supplies the derivation. This obstruction concerns one reward family, not all additive rewards.

\subsection{Constraints and deterministic costs}
For historical data we use a separate, fitted one-period target
\begin{equation}\label{eq:fitted}
 F(u)=m^\top u-\varphi u^\top\Sigma u-g(u),\quad u\in C,
\end{equation}
where $C $ is the simplex and $g(u)=\lambda\|u-p\|_1$ at pre-trade holdings $p $. Its optimum $u^*$ is not the true historical market optimum. For $a=m^\top u^*$ and $A=1+2\varphi a>0$, define
\begin{equation}
 Q_a(u)=m^\top u-\frac{\varphi}{A}u^\top Mu-\frac{g(u)}A.
\end{equation}
The identity
\begin{equation}\label{eq:aux}
 A Q_a(u)=F(u)-\varphi(m^\top u-a)^2+\varphi a^2
\end{equation}
proves that both objectives have the same maximiser. It is an application of the MVPI auxiliary~\cite{zhang2021}, not a new general algorithm. Correcting risk alone introduces the extra term $-2\varphi a\,g(u)$ and can change the cost preference.

For any anchor $a $ with $A>0$, an $\epsilon $-optimal $\widehat u $ for $Q_a $ satisfies
\begin{equation}\label{eq:anchor}
 F(u^*)-F(\widehat u)\le\varphi(m^\top u^*-a)^2+A\epsilon.
\end{equation}
This follows by comparing Eq.~\eqref{eq:aux} at $u^*$ and $\widehat u $. For convex $C,g $ and $\Sigma\succ0$, divide by $\varphi\lambda_{\min}(\Sigma)$ to bound squared allocation distance. The result includes deterministic borrowing and shorting charges, provided all costs are scaled. Choosing $a=m^\top p $ gives one existing MVPI update without supplying $u^*$; it is approximate rather than an exact calibration certificate.

\section{Certifying a learned equilibrium}
\subsection{A decision-relevant critic error}
Assume finite second moments, $\Sigma_t(z)\succ0$, and that future innovations are conditionally independent of $R_t $ given $Z_{t+1}$. Let $P_t $ denote the exogenous transition matrix. For a candidate policy $u $, the expected remaining terminal excess gain satisfies
\begin{equation}\label{eq:a}
 a_T=0,\qquad a_t=H_tm_t^\top u_t+P_ta_{t+1}.
\end{equation}
A one-step deviation to $v $ changes the terminal criterion through
\begin{equation}\label{eq:local}
 J_t(v)=H_tm_t^\top v-\varphi H_t^2v^\top\Sigma_tv
 -2\varphi H_t\Cov_t(R_t,a_{t+1}(Z'))^\top v+\mathrm{const}.
\end{equation}
The conditional independence and wealth-independent continuation assumptions justify the covariance reduction. They do not hold for arbitrary latent-state trading policies.

For any critic $c_t(z)$, define residuals and a critic-based action gradient by
\begin{align}
 e_t&=c_t-H_tm_t^\top u_t-P_tc_{t+1},\label{eq:residual}\\
 g_t&=H_tm_t-2\varphi H_t\Cov_t(R_t,c_{t+1}(Z'))
       -2\varphi H_t^2\Sigma_tu_t.\label{eq:gradient}
\end{align}
Use $\spn(f)=\max_z f(z)-\min_z f(z)$ and the Dobrushin coefficient
$\rho_t=\frac12\max_{z,w}\sum_j|P_t(z,j)-P_t(w,j)|$. Set
\begin{equation}\label{eq:span}
 S_T=\spn(c_T),\qquad S_t=\spn(e_t)+\rho_tS_{t+1}.
\end{equation}
No equilibrium policy appears in these quantities.

\begin{theorem}\label{th:audit}
Let $\mathcal E_t(z)=\sup_v[J_t(v)-J_t(u_t(z))]$. Under the preceding assumptions, with unrestricted deviations,
\begin{equation}\label{eq:audit}
 \sqrt{\mathcal E_t(z)}\le
 \frac{\|g_t(z)\|_{\Sigma_t(z)^{-1}}}{2\sqrt\varphi H_t}
 +\frac{\sqrt\varphi}{2}S_{t+1}.
\end{equation}
For a feasible policy on a fixed nonempty closed convex action set, replace the first term by $\sqrt{\mathcal E_t^c(z)}$, the maximal feasible improvement in Eq.~\eqref{eq:local} with $a_{t+1}$ replaced by $c_{t+1}$, assuming its maximum is attained. The same bound holds.
\end{theorem}
\begin{proof}
Write $\delta=c-a $. Equations~\eqref{eq:a}--\eqref{eq:residual} give $\delta_t=e_t+P_t\delta_{t+1}$. Stochastic-kernel span contraction implies $\spn(\delta_t)\le S_t $. The true action gradient is $d_t=g_t+2\varphi H_t\Cov_t(R_t,\delta_{t+1})$. Completing the square yields
$\mathcal E_t=\|d_t\|_{\Sigma_t^{-1}}^2/(4\varphi H_t^2)$.
For any scalar $D $, linear-projection residual variance is nonnegative, so
\[
\|\Cov(R,D)\|_{\Sigma^{-1}}^2\le\Var(D)\le\spn(D)^2/4.
\]
The triangle inequality proves Eq.~\eqref{eq:audit}.

For constraints, write $J=J_c+b^\top v+\mathrm{const}$, $b=2\varphi H\Cov(R,\delta)$ and $k=\varphi H^2$. If $v_c $ maximises $J_c $, strong concavity and its variational inequality imply
$J_c(v)-J_c(v_c)\le-k\|v-v_c\|_\Sigma^2$ and
$\|v_c-u\|_\Sigma\le\sqrt{\mathcal E^c/k}$. Thus
\[
 \mathcal E\le\mathcal E^c+\|b\|_{\Sigma^{-1}}\sqrt{\mathcal E^c/k}
       +\frac{\|b\|_{\Sigma^{-1}}^2}{4k}
 =\left(\sqrt{\mathcal E^c}+\frac{\|b\|_{\Sigma^{-1}}}{2\sqrt k}\right)^2.
\]
Apply the preceding covariance bound.
\end{proof}
The maximum certified $\mathcal E_t(z)$ bounds a one-step equilibrium deviation, not global-policy regret. Adding arbitrary time-dependent state-constant offsets to $c $ leaves the certificate unchanged. If $\rho_t\le\rho<1$ and residual spans are at most $\eta $, the critic contribution depends on a geometric sum, rather than an unweighted horizon sum. This uses established seminorm geometry~\cite{chen2025seminorm}. The vector result has no explicit dimension factor in the critic term; reliable covariance estimation remains dimension-dependent.

\subsection{Reward error, actor error and sampled audits}
If a learner follows a different action gradient $h_t $, split $g_t=h_t+(g_t-h_t)$ in Eq.~\eqref{eq:audit}. This gives nonnegative actor, reward and critic terms, respectively. It is an upper bound, not an additive decomposition of realised returns. A small neural parameter gradient is insufficient: a saturated actor may have a small parameter derivative despite a large action gradient. At a constraint boundary, use the feasible improvement in Theorem~\ref{th:audit} rather than an unconstrained gradient norm.

The exact-model certificate uses conditional moments. For scalar bounded returns, a separate independent-data construction estimates its gradient, variance and Bellman residuals using confidence bounds (Section~\ref{app:data}). It conditions on a policy and critic fixed before the audit and does not use the equilibrium policy. If the variance lower bound is nonpositive, it returns an uninformative infinite bound. These conditions rule out silently applying the guarantee to reused training data or a dependent historical time series.

\section{Learning and evaluation protocol}
\subsection{Continuation-aware sample learning}
Two independent conditional observations $(R_1,Z'_1),(R_2,Z'_2)$ yield the reward
\begin{equation}\label{eq:paired}
 L_c(v)=\frac{H(R_1+R_2)^\top v}{2}
 -\frac\varphi2\left[H(R_1-R_2)^\top v+c(Z'_1)-c(Z'_2)\right]^2.
\end{equation}
Its expected action gradient equals Eq.~\eqref{eq:gradient}. A backward mean critic learns from targets $HR^\top u+c(Z')$. This replaces oracle continuation by sampled policy evaluation, using established backward equilibrium-learning principles~\cite{lesmana2025}. Paired covariance estimation itself is standard.

\paragraph{Learner pseudocode (shared neural variant).}
Given conditional samples, $\varphi$, $T$ and a fixed update budget:
\begin{enumerate}
\setlength{\itemsep}{0pt}
\setlength{\parsep}{0pt}
\item Initialise actor $u_\theta$, critic $c_\psi$ and target $\bar c$, with zero terminal critics; freeze sample-based curvature weights.
\item Draw minibatches in each $(t,z)$ cell. Regress $c_\psi$ onto cell means of detached $HR^\top u_\theta+\bar c(Z')$.
\item Pair observations; with critic fixed, ascend curvature-weighted mean $L_{c_\psi}(u_\theta)$, Eq.~\eqref{eq:paired}, in $\theta$.
\item Update $\bar\psi\leftarrow .98\bar\psi+.02\psi$; repeat. The centred variant centres critic outputs and TD targets across regimes at each time. The myopic control zeros step~3's continuation difference.
\item Freeze policy and critic; audit independently (Section~\ref{app:data}). Equilibrium is used only for controlled evaluation.
\end{enumerate}

Our controlled market has two observed regimes, transition probabilities $(p,1-p)$ and $(1-p,p)$, conditional edge means
\[
 \begin{pmatrix}.09&-.05\\.04&-.02\end{pmatrix},
\]
and independent innovations $\pm.12$ with equal probabilities. We set $s=1.02$, $\varphi=1$, $X_0=1$, $p\in\{.5,.8,.95\}$ and $T\in\{4,12,24\}$. Even $p=.5$ is not an iid-return experiment: returns remain jointly related to successor regimes. Exact moments and equilibrium are restricted to evaluation.

The tabular study crosses 128/512/2,048/8,192 samples per time/state with twenty seeds and three methods: myopic paired variance, continuation-aware actor updates and exact sampled-objective optimisation (2,160 records). Forty actor steps use one-quarter estimated inverse curvature; zero variance is rejected.

The neural study uses 2,048 conditional samples per time/state, shared actor and critic networks with two 32-unit tanh layers, and inputs consisting of normalised time and regime one-hot encoding. Adam uses learning rate .003, target-critic averaging .02 and batch size 64 per time/state. We retain checkpoints at 250, 1,000 and 3,000 updates. The three variants use myopic reward, continuation-aware reward with an ordinary mean critic, or continuation-aware reward with state-centred critic outputs and TD targets. Hyperparameters were fixed before the five-seed pilot and retained when extending to twenty seeds. All methods share samples and initialisations. Actor losses use the same frozen, sample-estimated inverse-curvature scaling. Actions do not affect the market, and all sampled returns are observed; this generative-data setting is stronger than a single historical trajectory.

\subsection{A standard policy-optimisation comparison}
We additionally use clipped PPO~\cite{schulman2017}, Eq.~(7), via Stable-Baselines3 2.9.0. The two arms differ only in whether Eq.~\eqref{eq:paired} includes continuation covariance. Both use sample-only backward evaluation of the current deterministic policy, refreshed between rollouts; no exact equilibrium enters training. This is a structured PPO hybrid, not a reproduction of a published financial agent or a claim that PPO itself solves time inconsistency.

The protocol fixes $T=12$, $p=.8/.95$, seeds 0--9, 32,768 environment steps, actions in $[-3,3]$, and the same 2,048-sample conditional datasets. Each environment step draws a pair; the first successor advances the episode. Eight environments supply 512 observations per rollout, with ten epochs, batch size 128, learning rate $3\times10^{-4}$, clipping threshold .2, GAE parameter .95, discount 1, entropy coefficient 0, and separate 64-by-64 tanh actor/value networks. We evaluate the clipped deterministic deployment policy. Budgets are matched within PPO, not between PPO and direct-gradient learning.

\subsection{Historical allocation and common metrics}
Kenneth French's value-weighted industry returns and factors~\cite{french}, archived August 2026 vintage, provide 432 evaluation months from January 1990 to December 2025. The lagged window is 120 months; means are unshrunk excess-return means, $\varphi=5$ and L1-cost coefficient $\lambda=.001$. The data are revised research portfolios, not point-in-time investable funds. This experiment audits Eq.~\eqref{eq:fitted}; it is not a backtest of the new regime-based RL learners.

Seven covariance models (sample, fixed 10\% diagonal shrinkage, Ledoit--Wolf identity, constant correlation and nonlinear shrinkage, market factor and three Fama--French factors) are crossed with 10/30/49 industries. Within each model, rewards share target pre-trade holdings, moments and constraints. Cross-model comparisons use the diagonal-model target's holdings. Own-state paths determine realised performance. These controls separate covariance changes, reward changes and portfolio drift.

Controlled RL metrics are $D=(\sum_t\mathbb E_{u^{\eq}}\|u_t-u_t^{\eq}\|^2)^{1/2}$ under exact reference occupancy, starting in regime zero, and maximum one-step deviation gain. Initial terminal MV is secondary: equilibrium need not maximise the initial precommitment objective. Historical metrics are allocation distance, fitted regret, net excess-return MV, excess Sharpe and turnover. Paired neural uncertainty resamples seed-level averages across the nine fixed markets; historical uncertainty uses shared circular twelve-month blocks. Neither treats related market configurations as independent new datasets.

\section{Results}
\subsection{Learning error and residual certification}
The tabular continuation-aware learner's average distance decreases from 1.436 at 128 samples to 0.1649 at 8,192, while myopic distance decreases from 1.497 to 0.5871. Corresponding maximum deviation gains at 8,192 are 0.000273 and 0.003319. More data reduces sampling error but does not remove an omitted continuation hedge. The maximum difference in distance between forty actor steps and exact optimisation of the same sampled objective is $5.24\times10^{-5}$.

\begin{figure*}[t]
\centering\includegraphics[width=\textwidth]{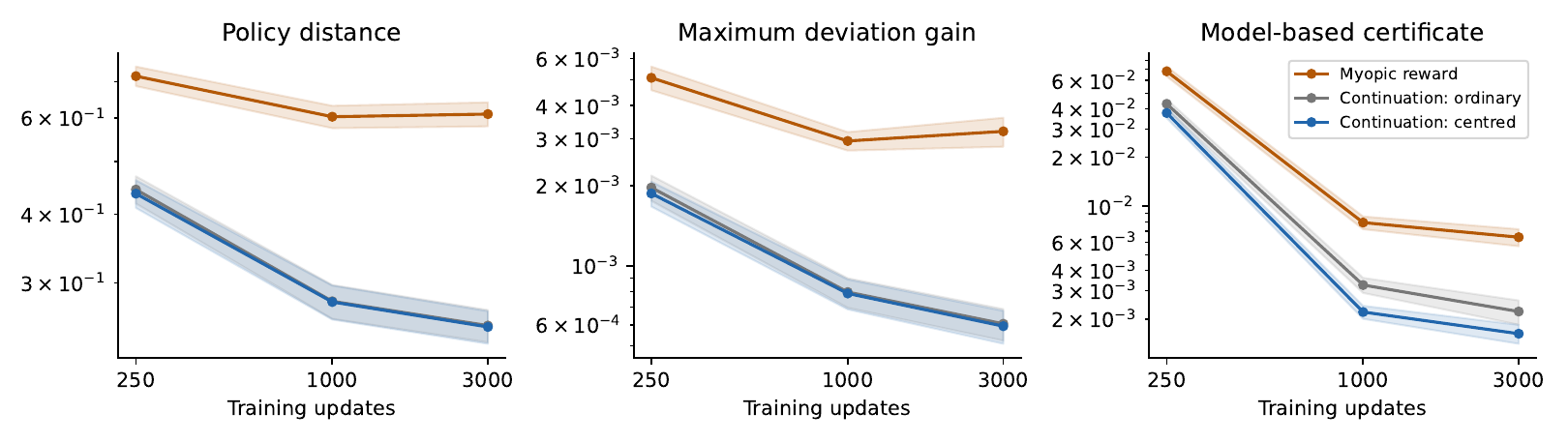}
\caption{Neural learning across twenty seeds and nine markets. Markets are averaged within seed; bands show descriptive mean $\pm1.96$ standard errors, not simultaneous intervals. The certificate audits policy-relevant errors.}
\Description{Three learning curves compare policy distance, true deviation gain and the model-based residual certificate for myopic, ordinary continuation and centred continuation learners.}
\label{fig:rl}
\end{figure*}

Table~\ref{tab:rl} and Figure~\ref{fig:rl} report the neural comparison. At the fixed final budget, the centred continuation-aware variant reduces mean distance by \DistanceReduction\% and deviation gain by \DeviationReduction\% relative to the matched myopic network. The paired distance difference is $-0.361$ (95\% seed-bootstrap interval $[-0.390,-0.329]$). Its additional distance improvement over the ordinary continuation critic is \CentreReduction\%; most of the improvement comes from the continuation term. The centring effect on distance is small: $-0.00143$, with paired interval $[-0.00253,-0.00030]$. Centring reduces the mean certificate by 27.1\%; it has a larger effect on auditing than on the deployed policy. We report all seeds and do not select checkpoints using oracle quality.

\begin{table}[t]
\centering
\caption{Controlled RL: A averages nine markets and twenty seeds (3,000 updates); B averages two markets and ten seeds (32,768 PPO steps). $D $: policy distance; $\mathcal E $: maximum feasible deviation gain. Compute is matched within panels only. Bold: continuation-aware method.}
\label{tab:rl}
\begin{tabular}{lrrr}
\toprule
Method & $D$ & $\mathcal E$ & Bound \\
\midrule
\multicolumn{4}{l}{A. Shared neural actor--critic} \\
Myopic reward & 0.6102 & 0.00320 & 0.00642 \\
Continuation, ordinary & 0.2507 & 0.00061 & 0.00223 \\
\textbf{Continuation, centred (ours)} & \textbf{0.2493} & \textbf{0.00059} & \textbf{0.00163} \\
\midrule
\multicolumn{4}{l}{B. Clipped PPO; separate matched comparison} \\
PPO, myopic & 1.7280 & 0.02349 & 0.03279 \\
\textbf{PPO, continuation (ours)} & \textbf{1.1525} & \textbf{0.01180} & \textbf{0.01723} \\
\bottomrule
\end{tabular}

\end{table}

\PpoText\ The PPO result supports a reward intervention within a fixed optimiser; it does not establish algorithmic superiority over the recent equilibrium-RL frameworks in Section~2. Their continuous-time problems and model access differ from this benchmark.

A false-convergence construction has zero critic-based action gradients but true deviation gain 0.10263, bounded by 0.89674. State-constant critic offsets change bounds by less than $1.7\times10^{-15}$ across twenty cases. Independent terminal-moment tests check random policies and active constraints; vector projection checks do not constitute high-dimensional neural experiments.

\subsection{Auditing with independent samples}
\begin{figure}[t]
\centering\includegraphics[width=\columnwidth]{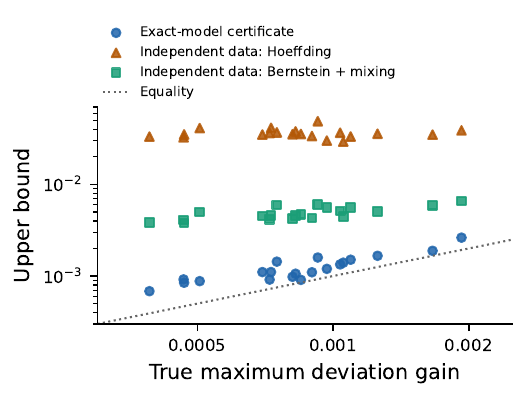}
\caption{Independent-data audits of twenty fixed policies, each using 786,432 observations. Refinement tightens Hoeffding bounds. Coverage follows from the stated assumptions, not the plotted outcomes.}
\Description{Logarithmic scatter plot of true deviation gain against exact-model, Hoeffding and refined independent-data upper bounds. All displayed bounds exceed the actual gains.}
\label{fig:audit}
\end{figure}

Twenty fixed tabular policies in the $T=12,p=.8$ market receive 32,768 independent audit observations per time/state, or 786,432 each. Figure~\ref{fig:audit} compares bounds for each policy. Mean actual maximum deviation is 0.000902 and the exact-model bound is 0.001257. A Hoeffding-only observable bound averages 0.036141. Empirical Bernstein bounds with confidence-bounded transition mixing reduce this to 0.004911. Every policy is covered by each construction. Coverage in these twenty cases is a numerical check, not the source of the confidence guarantee.

The Bernstein and mixing refinements were introduced after inspecting the Hoeffding bounds, so we report them as a post hoc extension. Each construction has its own 95\% simultaneous-over-states guarantee for a fixed policy; no joint 95\% guarantee across all policies or constructions is claimed. Large audit requirements and abstention when variance cannot be certified are substantive properties of the method.

\subsection{Financial inputs and implemented preferences}
Table~\ref{tab:historical} compares published reward formulations and covariance models on historical data. Its cross-estimator distances are changes in fitted models, not pure reward errors. The restricted Habibnia--Soltanzadeh comparator uses long-only unit-capital assetwise PnL, penalty 25 and cost multiplier $25/5$; SAC, shorting and lending are not replicated. Its objective depends on marginal downside payoffs and can prefer allocations different from mean--variance diversification.

\begin{table*}[t]
\centering
\caption{Ten industries, 1990--2025. Distances/gaps use common states and the fixed-diagonal reference; covariance rows change the model. Realised metrics use own-state paths. Sources are chronological; 2026 denotes implemented controls. Published formulations are adapted, not full agent replications. Bold identifies our construction, not investment superiority.}
\label{tab:historical}
\begin{tabular}{@{}p{.235\textwidth}p{.25\textwidth}rrrr@{}}\toprule
Method & Source / role & \shortstack{Allocation\\distance} & \shortstack{Reference gap\\(bp)} & \shortstack{Excess\\Sharpe} & \shortstack{Realised MV\\(bp)}\\\midrule
Fitted MV reference & Markowitz (1952) & 0.00000 & 0.0000 & 0.570 & $-$8.497 \\
Minimum variance & Markowitz (1952) & 0.23843 & 9.0252 & 0.645 & 5.274 \\
MV, LW identity & Ledoit--Wolf (2004) & 0.00538 & 0.0055 & 0.572 & $-$8.442 \\
MV, LW correlation & Ledoit--Wolf (2004) & 0.02196 & 0.1158 & 0.547 & $-$11.917 \\
MV, market factor & Fan et al. (2008) & 0.05711 & 1.0487 & 0.585 & $-$4.738 \\
MV, three factors & Fan et al. (2008) & 0.04881 & 0.6407 & 0.555 & $-$8.532 \\
Equal weight & DeMiguel et al. (2009) & 0.43495 & 46.2626 & 0.622 & $-$11.893 \\
MV, LW nonlinear & Ledoit--Wolf (2017) & 0.01119 & 0.0209 & 0.559 & $-$9.578 \\
Raw second moment & Kolm--Ritter (2019) & 0.00843 & 0.0148 & 0.573 & $-$7.182 \\
MVPI adaptation & Zhang et al. (2021) & 0.00000 & 0.0000 & 0.570 & $-$8.497 \\
One MVPI step & Zhang et al. (2021) & 0.00014 & 0.0000 & 0.570 & $-$8.442 \\
Assetwise PnL, restricted & Habibnia--Soltanzadeh (2024) & 0.79645 & 40.2554 & 0.449 & $-$23.619 \\
Log minus variance & Jiang et al. (2025) & 0.01286 & 0.0287 & 0.572 & $-$7.123 \\
MV, sample covariance & Sample control (2026) & 0.01449 & 0.0356 & 0.555 & $-$10.114 \\
Log utility & Log-utility control (2026) & 0.54848 & 39.7855 & 0.344 & $-$133.295 \\
Unconstrained scalar & Analytical control (2026) & 0.00540 & 0.0043 & 0.568 & $-$8.882 \\
Risk-only calibration & Calibration ablation (2026) & 0.00510 & 0.0033 & 0.571 & $-$8.365 \\
\midrule
\textbf{Joint calibration (ours)} & \textbf{This study (2026)} & \textbf{0.00000} & \textbf{0.0000} & \textbf{0.570} & \textbf{$-$8.497} \\
\bottomrule\end{tabular}

\par\smallskip
\parbox{\textwidth}{\footnotesize Sources: Markowitz~\cite{markowitz1952}; Ledoit--Wolf~\cite{ledoit2004identity,ledoit2004correlation}; Fan et al.~\cite{fan2008}; DeMiguel et al.~\cite{demiguel2009}; Ledoit--Wolf~\cite{ledoit2017}; Kolm--Ritter~\cite{kolm2019}; Zhang et al.~\cite{zhang2021}; Habibnia--Soltanzadeh~\cite{habibnia2024}; Jiang et al.~\cite{jiang2025}.}
\end{table*}

Across all twenty-one covariance/universe cases, joint calibration has mean distance below $1.60\times10^{-7}$ from its own fitted reference. Raw distances range 0.00827--0.02249; transferring the unconstrained scalar formula worsens distance and fitted regret in all fourteen larger-universe cases. One existing MVPI update initialised at holdings reduces mean distance by 97.8--98.6\%, without supplying the fitted optimum. These are calibration results, not a new MVPI algorithm.

Dimension increases $N/L $ from 0.083 to 0.250 and 0.408, with 65/495/1,274 distinct mean/covariance entries. The classifications also change the opportunity set, so this is not a causal experiment in dimension alone. DeMiguel et al.'s~\cite{demiguel2009} 3,000/6,000-month examples are model-dependent illustrations, not universal requirements. Simple controls and shrinkage remain essential.

Minimum variance has the highest realised MV among quadratic comparators in all twenty-one cases, while joint calibration remains below raw. All eighteen cross-covariance target comparisons have paired 95\% intervals containing zero, including simultaneous path-bootstrap bands. Improved alignment therefore does not establish improved historical investment performance. In 2,160 known-covariance fitted problems using the six estimators other than nonlinear shrinkage, factor structure helps when its residual assumptions hold; misspecification can worsen population regret while leaving exact reward calibration intact. A separate fifty-state borrowing/shorting test gives joint distance $1.51\times10^{-7}$, versus 0.00997 when financing charges are not scaled. 

\begin{figure*}[t]
\centering\includegraphics[width=\textwidth]{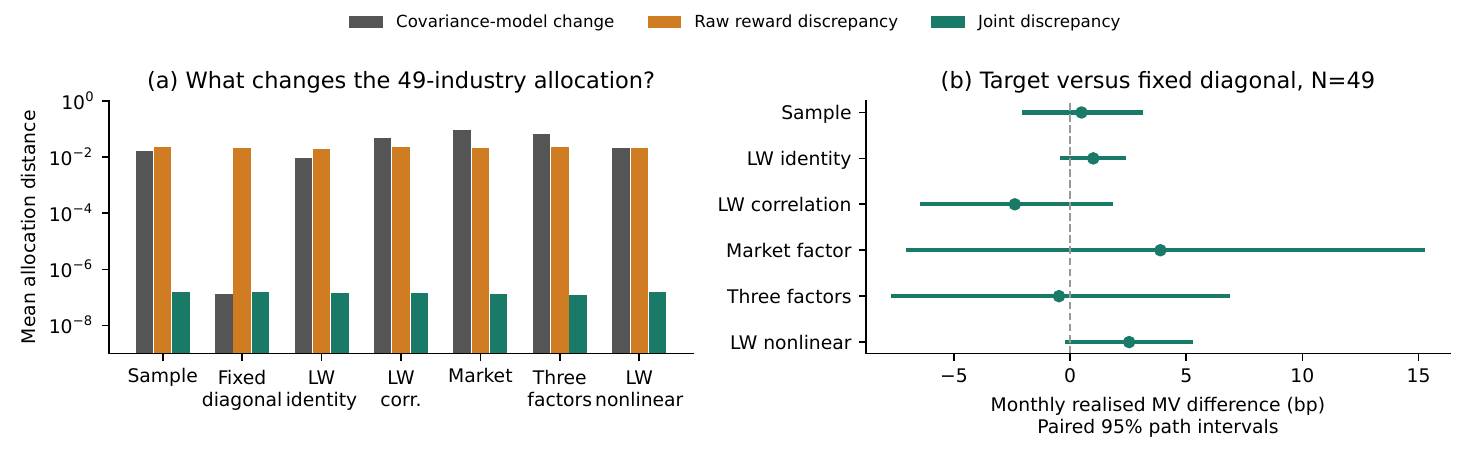}
\caption{49-industry covariance and reward interventions. Joint calibration preserves fitted preferences, without identifying true covariance or establishing superior own-state performance.}
\Description{Historical covariance comparisons distinguish policy alignment within each model from realised investment performance.}
\label{fig:cov}
\end{figure*}

\section{Financial interpretation and validation}
A financial explanation should distinguish changed opportunities (covariance), changed preferences (reward), and implementation error (actor or critic). In the 49-industry market-factor case, changing covariance moves the target by 0.09392; within-model raw and joint reward discrepancies are 0.02174 and $1.36\times10^{-7}$. Combining them into one error score would obscure which intervention changed the decision.

Nonlinear shrinkage strengthens the estimation control. Ledoit and Wolf~\cite{ledoit2017}, Table~8, caution against transferring unconstrained performance rankings to long-only portfolios. With nonlinear shrinkage, joint allocation distances are $9.40\times10^{-9}$, $8.67\times10^{-8}$ and $1.60\times10^{-7}$ for 10/30/49 industries. Its 49-industry target improves realised MV over fixed diagonal shrinkage by 2.54 monthly basis points, but the paired 95\% interval $[-0.15,5.20]$ includes zero. 

The learning audit explains whether changing an action could improve the stated objective. State-constant critic errors are harmless; state-dependent errors can conceal non-equilibrium decisions. Feasible deviations respect boundary optima. This is decision-level validation, not attribution of hidden neural representations.

\paragraph{A worked validation decision.}
Let a mandate tolerate at most $\tau=0.001$ in conditional terminal mean--variance improvement from any one-action change: ten basis points in the experiment's normalised utility units, not ten basis points of realised return. Require a 95\% simultaneous-over-time/state upper bound $B\le\tau$. For the first saved audit policy (seed~0), 786,432 fresh observations give $B=0.004992$ using Bernstein bounds and estimated mixing. The decision is \emph{inconclusive}: the bound exceeds the tolerance, although the simulator's true maximum gain, 0.000506, is below it. A separate mandate allowing $\tau=0.01$ would pass the same bound; Hoeffding alone ($B=0.041214$) would certify neither. These are illustrative tolerances applied retrospectively, not preregistered acceptance results. Operational use must fix the tolerance, policy and audit construction before inspecting audit outcomes; an infinite bound also requires abstention.

\paragraph{Scope of the evidence.}
The sequential experiments establish the correction's effect in a small observed-state market with conditional sampling. The multiasset historical study instead establishes alignment with a fitted one-period objective; it does not validate the sequential learner or its independent-data guarantee on market histories. A realistic sequential multiasset test remains open. Unknown regimes, market impact, wealth-dependent policies and transaction-cost states require additional continuation analysis, while dependent audit data require new statistical guarantees.

Kolm and Ritter~\cite{kolm2025}, Exhibit~2 and pp.~90--91, distinguish reward specification, learning and decision quality. Our audit connects these stages within model-risk review, taking the mandate as given. Direct optimisation remains appropriate for tractable one-period allocation; sequential learning additionally needs an account of remaining decision error.

\section{Independent-data certificate}\label{app:data}
Fix a scalar policy $u $ and critic $c $ independently of audit data, and assume known support $|R|\le B $. At every $(t,z)$ obtain $N=2n $ independent conditional observations. Pair them and differentiate Eq.~\eqref{eq:paired}; each gradient sample has absolute bound
\[
 M_g=HB+2\varphi HB\spn(c_{t+1})+4\varphi H^2B^2|u|.
\]
The pair variable $(R_1-R_2)^2/2$ estimates conditional variance and lies in $[0,2B^2]$. Bellman targets have range at most $2HB|u|+\spn(c_{t+1})$. Hoeffding radii with failure probability $q=\alpha/(3TK)$ give a gradient upper norm, variance lower bound and residual intervals $[l_z,h_z]$. Then $\spn(e_t)\le\max_z h_z-\min_z l_z $; use $\rho_t\le1$ in Eq.~\eqref{eq:span}. Substitute these bounds in Eq.~\eqref{eq:audit}. The result covers all time/state cells with probability at least $1-\alpha $, conditional on training.

For a variable of known range width $W $ and sample size $n $, Maurer--Pontil~\cite{maurer2009}, Theorem~4, gives the two-sided radius
\[
 \sqrt{\frac{2\widehat V\log(4/q)}n}
 +\frac{7W\log(4/q)}{3(n-1)},
\]
where $\widehat V $ uses denominator $n-1$. Confidence intervals for each transition entry additionally give $\rho_t\le\min(1,\widehat\rho_t+K\eta)$, where $\eta $ bounds each entry error. Include $TK^2$ additional events in the union bound. The construction returns infinity if any variance lower bound is nonpositive. Adaptive policy selection on the audit set, dependent observations or unknown support require additional analysis; none is assumed here. The scalar implementation bounds unrestricted deviations, which also upper-bound constrained ones.

\section{Raw-reward discrepancy}\label{app:raw}
For constant independent moments with $m\ne0$, the quadratic Bellman solution has $u_t=\alpha_tx+\beta_t $, with terminal $p_T=q_T=0$ and $g_t=\kappa(s-1)-sp_{t+1}$:
\begin{align*}
\alpha_t&=-dg_t/(\kappa-p_{t+1}),\\
\beta_t&=d(1+q_{t+1})/[2(\kappa-p_{t+1})],\\
p_t&=[\kappa p_{t+1}-(1-\nu)g_t^2]/(\kappa-p_{t+1}),\\
1+q_t&=(1+q_{t+1})[s-\nu g_t/(\kappa-p_{t+1})].
\end{align*}
Induction gives $p_t\le0$, hence global concave action maximisation. Put $e=s-1$ and $h=(2\kappa)^{-1}$. Because $p_t/\kappa,q_t $ are independent of $\kappa $, the wealth recursion is affine in $h $ and $D^2$ is exactly quadratic in $h $, analytic near $e=0$. At zero it is $T\|d\|^2(h-b)^2$, so its positive minimiser is analytic nearby.

Writing $\kappa=\varphi(1-\nu)(1+we)$, the leading action error is $de[-X_t+b((T-t-1)(2-\nu)-w)]$. At $e=0$, mean wealth is $x_0+t\nu b $ and variance $tb^2\nu(1-\nu)$. Minimisation gives $w=(T-1)(1-\nu)-x_0/b $; summing centred mean squares $b^2(T-1-2t)^2$ and wealth variances yields Eq.~\eqref{eq:raw}. This is a fixed-horizon expansion, not a uniform joint limit. For $s\ne1$, exact matching is impossible: matching initially creates positive subsequent wealth variance, whereas the final raw policy has nonzero wealth slope $-d(s-1)$ and equilibrium is constant.

\section{Reproduction and numerical checks}\label{app:repro}
Conditional-sample generators use seeds 0--19, also controlling neural weights and minibatches; PPO uses seeds 0--9. Both PPO arms share data access and refresh frequency; continuation is fitted from training outcomes, not the analytic reference. The pilot, twenty-seed extension and PPO experiment are archived separately. Bootstrap intervals use 10,000 resamples of seed averages; all final-budget runs are retained.

Historical identity shrinkage uses centred $Z $, $S_0=Z^\top Z/L $ and target $\operatorname{tr}(S_0)I/N $, with intensity from the empirical fourth moment; final covariance is converted to denominator $L-1$. Constant-correlation shrinkage uses the plug-in formulas of Ledoit and Wolf~\cite{ledoit2004correlation}. Factor models use intercept OLS, market/SMB/HML and diagonal residual variances with denominator $L-K-1$. This convention adapts Fan et al.~\cite{fan2008}. Nonlinear shrinkage uses the covariance branch of QuEST v027 through a pinned Python port, with trace-constrained SLSQP fitting, tolerance $10^{-9}$, and effective sample size $L-1$ after demeaning. No additional rescaling is applied. This implements the estimator, not the original CRSP backtest. Missing and duplicate months are rejected. Holdings drift with gross asset returns; costs are charged separately at the stated first-order rate.

Historical uncertainty uses 1,000 shared circular twelve-month blocks, seed 8128, conditional on realised strategy paths; simultaneous covariance bands use the maximum centred deviation across eighteen comparisons. Quadratic optimisation is checked independently by a first-order linear-programme certificate in original objective units, with accepted gap at most $2\times10^{-7}$. Reconstruction verifies 67,392 historical decisions, 2,160 known-covariance fitted problems and 144 anchor checks. RL tests independently evaluate terminal moments, gradient/covariance identities, critic offsets, active boundaries and audit support. 

\section{Conclusion}
The residual certificate connects learning errors to financial decision quality, respecting critic-offset invariance and convex constraints. Controlled actor--critic and PPO comparisons test continuation correction without an oracle target. Historical experiments separately test fitted reward alignment. Together they support validation that distinguishes objective fidelity from investment performance and explicitly abstains when audit evidence is insufficient.

\section*{Declarations}
\paragraph{Funding and computational resources.}
This study received no project-specific research funding. No computing resources from Virginia Tech or the authors' other affiliated universities were used in this study.

\paragraph{Code and data availability.}
The simulation experiments use data generated by the accompanying code. The historical analysis uses publicly available monthly value-weighted US industry portfolio returns and factors from Kenneth French's Data Library~\cite{french}, using the archived August 2026 vintage. Replication code, archived input data, experiment outputs and provenance records are available at \url{https://github.com/DataismLab/Reward-alignment}.

\bibliographystyle{ACM-Reference-Format}
\bibliography{references}

\end{document}